\documentclass[epj]{svjour}
\usepackage{graphics}
\usepackage[dvips]{graphicx}
\let\Otemize =\itemize
\let\Onumerate =\enumerate
\let\Oescription =\description
\def\Nospacing{\itemsep=0pt\topsep=0pt\partopsep=0pt\parskip=0pt\parsep=0pt}
\def\Topspac{\vspace{-0.5\baselineskip}}
\def\Botspac{\vspace{-0.2\baselineskip}}
\newenvironment{Itemize}{\Topspac\Otemize\Nospacing}{\endlist\Botspac}
\newenvironment{Enumerate}{\Topspac\Onumerate\Nospacing}{\endlist\Botspac}

\newcommand{\lsim}{\,{\buildrel < \over {_\sim}}\,}
\newcommand{\gsim}{\,{\buildrel > \over {_\sim}}\,}
\newcommand{\sqrtsNN}{\sqrt{s_{\scriptscriptstyle{{\rm NN}}}}}
\newcommand{\av}[1]{\left\langle #1 \right\rangle}

\newcommand{\gev}{\mathrm{GeV}}
\newcommand{\tev}{\mathrm{TeV}}
\newcommand{\fm}{\mathrm{fm}}

\newcommand{\cm}{\mathrm{cm}}

\newcommand{\mum}{\mathrm{\mu m}}

\newcommand{\PbPb}{\mbox{Pb--Pb}}

\newcommand{\RAA}{R_{\rm AA}}
\newcommand{\RDh}{R_{{\rm D}/h}}
\newcommand{\pt}{p_{\rm t}}
\renewcommand{\d}{{\rm d}}
\newcommand{\dEdx}{{\rm d}E/{\rm d}x}
\newcommand{\dNdy}{{\rm d}N_{\rm ch}/{\rm d}y}

\newcommand{\Dz}{\mbox{$\mathrm {D^0}$}}
\newcommand{\DtoKpi}{\mbox{${\rm D^0\to K^-\pi^+}$}}

\newcommand{\Jpsi} {\mbox{J\kern-0.05em /\kern-0.05em$\psi$}\xspace}

\begin{document}
\title{ALICE perspectives for the study of charm and beauty 
\mbox{energy loss at the LHC}}
\titlerunning{ALICE perspectives for the study of c and b energy loss 
at the LHC}
\author{A.~Dainese for the ALICE Collaboration% etc
}                     % Do not remove
\institute{INFN -- LNL, Legnaro (Padova), Italy}
\date{Received: \today}
\abstract{
At LHC energy, heavy quarks will be abundantly produced
and the design of the ALICE detector will allow us to
study their production using several channels.
The expected heavy-quark in-medium energy loss in nucleus-nucleus
collisions at the LHC is calculated within a model, that is
compared to the available heavy-quark quenching measurements at RHIC.
The nuclear
modification factors and heavy-to-light ratios
of charm and beauty mesons are considered.
The capability of the ALICE experiment for addressing
this phenomenology is discussed.
 % end of PACS codes%
} %end of abstract

\maketitle
\section{Introduction}
\label{intro}

The ALICE experiment~\cite{alicePPR1,alicePPR2} 
will study Pb--Pb
collisions at the LHC, with a centre-of-mass energy $\sqrtsNN=5.5~\tev$, 
in order to investigate the properties of QCD matter at energy densities of 
up to several hundred times the density of atomic nuclei. Under 
these conditions
a deconfined state of quarks and gluons is expected to be formed.

The measurement of open charm and open beauty production allows to 
investigate the mechanisms of heavy-quark production and propagation
in the hot and dense medium formed in the collision.
In particular, medium-induced partonic energy loss of heavy quarks 
has recently become one of the most exciting and disputed issues 
within high-energy heavy-ion physics,
after the observation at RHIC of a large 
suppression in the production of high-transverse-momentum 
electrons from heavy-flavour decays.
In Section~\ref{eloss} we describe these experimental results 
and we compare them to a model implementation of a particular 
energy loss calculation. Within this model we then obtain predictions
for relevant charm and beauty energy loss observables at LHC energy.
After providing, in Section~\ref{exp}, 
a general overview of the heavy-flavour capabilities
of the ALICE detector, we present the expected sensitivity in the measurement
of energy loss effects, in Section~\ref{sensitivity}. 

\section{Heavy-quark energy loss from RHIC to LHC} 
\label{eloss}

Believed to be the main origin of the jet quenching effects 
observed~\cite{dunlop} in 
nucleus--nucleus collisions at RHIC energy 
$\sqrtsNN=62$--$200~\gev$, 
parton energy loss via gluon-radiation is expected to depend
on the properties (gluon density and volume) of the medium
and on the properties (colour charge and mass) of the `probe' 
parton.
Gluons would lose more energy than quarks due to the stronger 
colour coupling.   
In addition, charm and beauty quarks are 
qualitatively different probes with respect to
 light partons, since their energy loss
is expected to be reduced, as a consequence of a mass-dependent restriction 
in the phase-space
into which gluon radiation can 
occur~\cite{dokshitzerkharzeev,aswmassive,djordjevic,zhang}.

Quenching effects for heavy quarks can be estimated by supplementing
perturbative QCD calculations of the baseline $\pt$ distributions 
with in-medium energy loss. Here, we consider the particular 
radiative energy loss calculation that is implemented in the 
BDMPS formalism~\cite{bdmps}.
The energy loss probability distributions (quenching weights)
were computed for light quarks and gluons in~\cite{sw} and 
for heavy quarks in~\cite{adsw}. They
depend on the medium
transport coefficient $\hat{q}$, the average transverse momentum 
squared transferred from probe parton to the scattering centres in the 
medium per unit mean free path, 
and on the in-medium path length $L$ of the probe parton. 
The collision geometry is included by evaluating 
$\hat{q}$ and $L$ on a parton-by-parton level, using a Glauber-model based 
description of the local $\hat{q}$ profile in the transverse 
plane~\cite{pqm}. The parton-averaged $\av{\hat{q}}$ value 
(hereafter called $\hat{q}$ for brevity) 
is chosen in order to reproduce the factor 4--5 suppression 
measured for the nuclear modification factor
\begin{equation}      
  \label{eq:raa}
  R_{\rm AA}(\pt) = 
  \frac{1}{\av{N_{\rm coll}}} \times
  \frac{{\rm d}^2 N_{\rm AA}/{\rm d}\pt{\rm d}\eta}
       {{\rm d}^2 N_{\rm pp}/{\rm d}\pt{\rm d}\eta}  
\end{equation}
of light-flavour particles in central 
\mbox{Au--Au} collisions at $\sqrtsNN=200~\gev$.
The range favoured by the data is 
$\hat{q}=4$--$14~\gev^2/\fm$~\cite{constantin,panic}. 

\begin{figure}[!t]
  \includegraphics[width=0.45\textwidth]{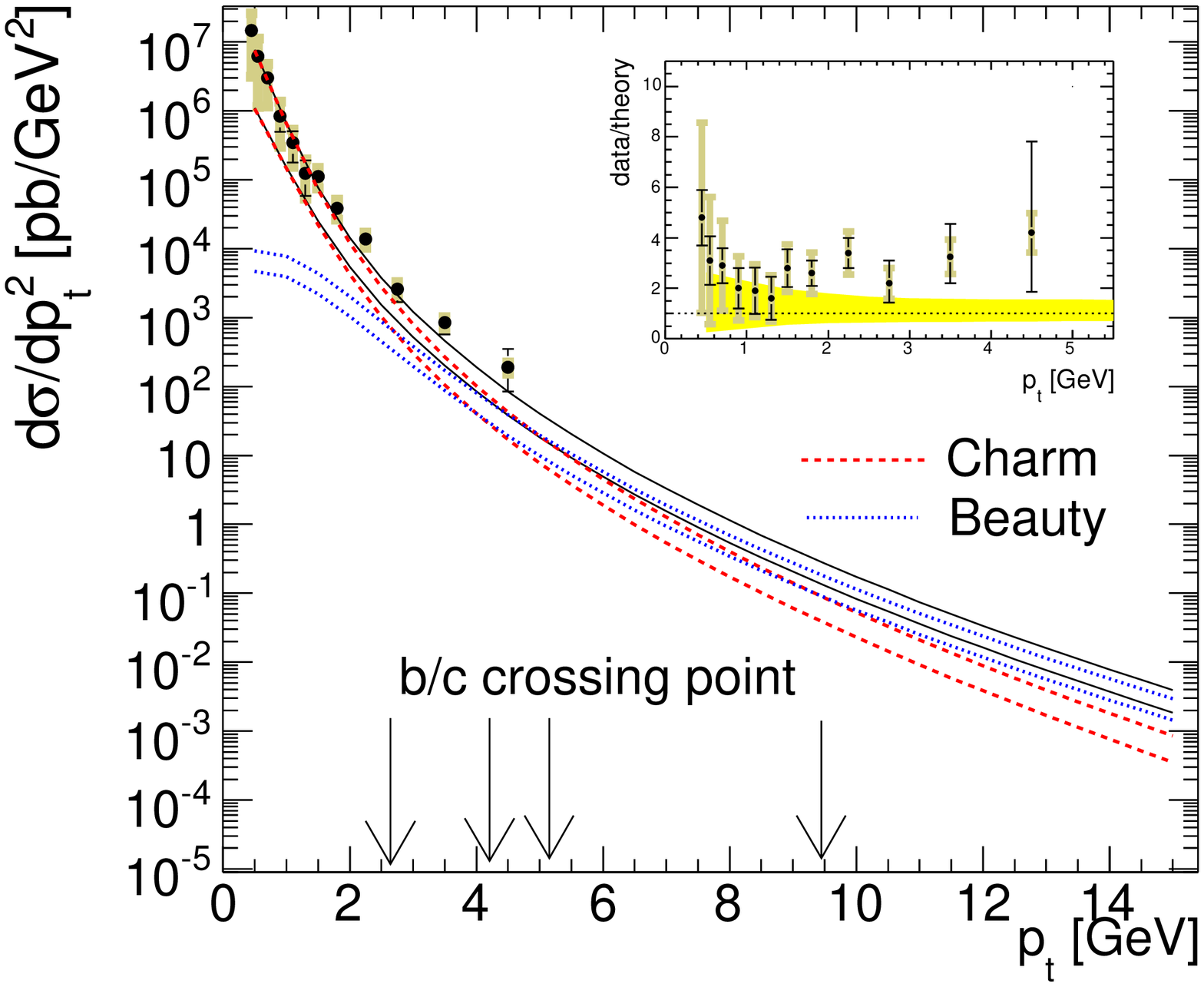}
  \includegraphics[angle=-90,width=0.45\textwidth]{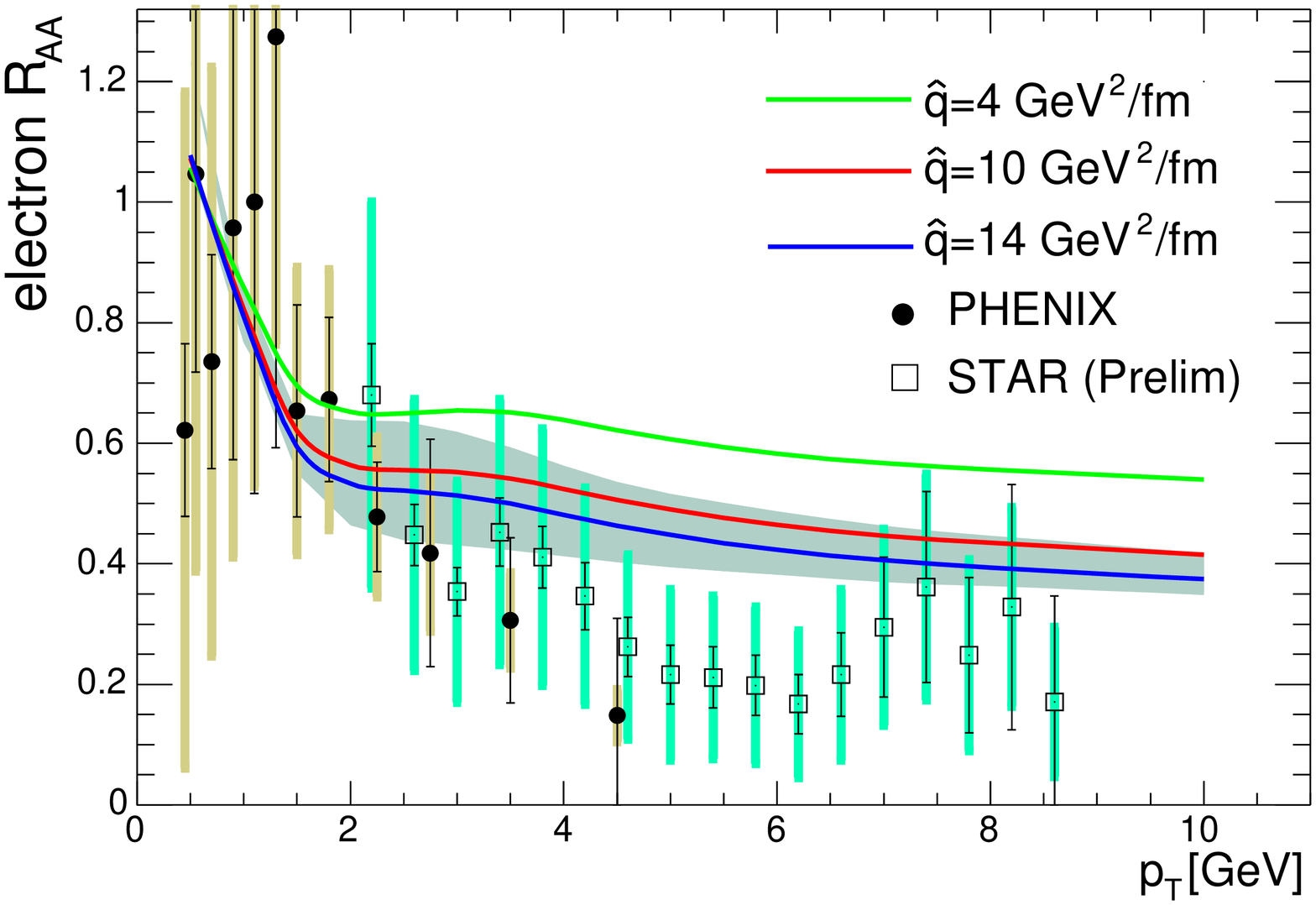}
 \caption{From~\cite{acdsw}.
          Upper panel: comparison of the FONLL calculation of 
          single inclusive electrons~\cite{Cacciari:2005rk} to 
          data from~\cite{Adler:2005fy} 
          pp collisions at $\sqrt{s} = 200~\gev$. 
          Upper and lower lines are estimates of theoretical uncertainties, 
          obtained by varying scales and masses. 
          Lower panel: $\RAA$ of electrons in 
          central Au--Au collisions~\cite{phenixe,stare}.
          Curves indicate the suppression for different opacities of the 
          produced matter.
          The shaded band indicates the theoretical uncertainty of the
          perturbative baseline for $\hat{q}=14~{\rm GeV}^2/{\rm fm}$.}
 \label{fig:rhic}
\end{figure}

Heavy-quark energy loss is presently studied at RHIC using
measurements of the nuclear modification factor $\RAA$ 
of `non-photonic' ($\gamma$-conversion- and $\pi^0$-Dalitz-subtracted) 
single electrons. 
Since this is an inclusive measurement, with charm decays dominating at low
$\pt$ and beauty decays dominating at high $\pt$, the comparison with 
mass-dependent 
energy loss predictions should rely on a solid and data-validated 
pp baseline. Such baseline is still lacking at the moment.
The state-of-the-art perturbative predictions (FONLL), usually employed
as a baseline, indicate that, in pp collisions, 
charm decays dominate the electron 
$\pt$ spectrum up to about 5~GeV~\cite{acdsw}. 
However, there is a large perturbative uncertainty on the
position in $\pt$ of the c-decay/b-decay 
crossing point: depending on the choice of the factorization 
and renormalization scales this position 
can vary from 3 to 9~GeV~\cite{acdsw}, as illustrated in Fig.~\ref{fig:rhic}
(upper panel).
In addition, as shown in the insert, the calculation underpredicts 
the non-photonic electron spectrum measured in pp 
collisions~\cite{Adler:2005fy}. 

The most recent data by PHENIX~\cite{phenixe} 
and STAR~\cite{stare} on the nuclear modification factor 
$\RAA^{\rm e}$
of non-photonic electrons in central \mbox{Au--Au} collisions at 
$\sqrtsNN=200~\gev$ are shown in Fig.~\ref{fig:rhic} (lower panel).
The theoretical expectation is superimposed to the data, 
with the uncertainty on the medium density (curves for 
$\hat{q}=4$, 10, $14~\gev^2/\fm$) and 
the perturbative uncertainty, obtained by varying the values of the
scales and of the c and b quark masses (shaded band associated to 
the $\hat{q}=14~\gev^2/\fm$ curve)~\cite{acdsw}.
The calculation tends to overpredict the measured $\RAA$.
It has recently been argued~\cite{djordjelastic} that parton energy 
loss would have a significant collisional contribution, 
comparable to the radiative one. Although the quantitative relevance of 
the collisional contribution is still debated~\cite{wangelastic},
the effect has been included in heavy-quark 
energy loss calculations~\cite{djordjelastic}.
Yet, the large suppression measured for $\RAA^{\rm e}$ can not be well 
reproduced~\cite{djordjelastic}.

It is important to note that, in general, the perturbative uncertainty in 
calculating the partonic baseline spectrum is comparable to the 
model-intrinsic uncertainty in determining $\hat{q}$. Thus,
the strongest limitation to the sensitivity in the
theory-data comparison comes from the inability 
of the RHIC experiments, in their present detector setup,
to disentangle the 
charm and beauty contributions to single electrons.

\begin{figure}[!t]
\includegraphics[width=0.5\textwidth]{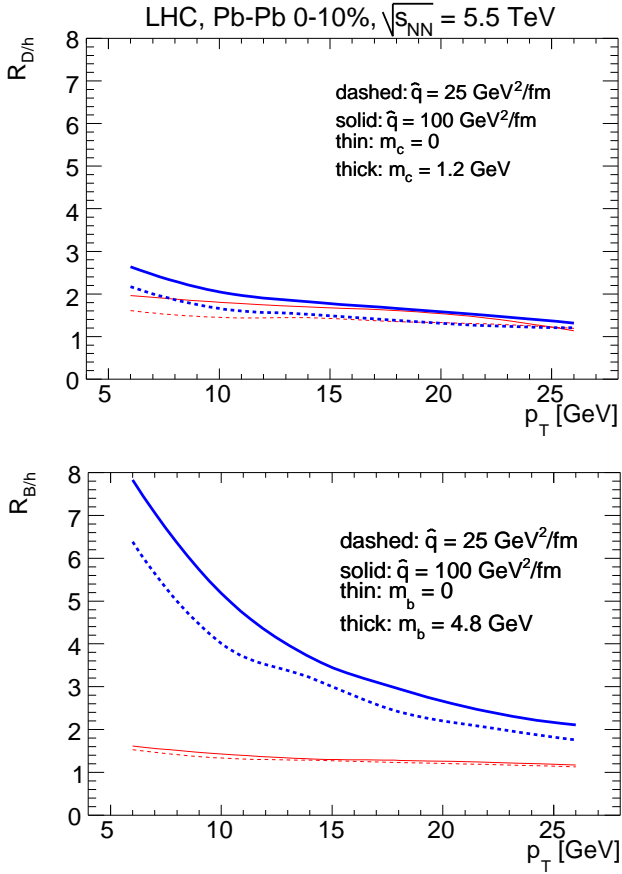}
\caption{Heavy-to-light ratios for D (upper panel) 
         and B (lower panel) mesons 
         for the case of realistic heavy-quark masses 
         and for a case
         study in which the quark mass dependence of parton energy 
         loss is neglected~\cite{adsw}.}
\label{fig:lhc}
\end{figure}

Heavy quarks will be produced with large cross sections at LHC energy 
and the experiments will be equipped with detectors optimized for the
separation of charm and beauty decay vertices. Thus, it will possible
to carry out a direct comparison of the attenuation of light-flavour
hadrons, D mesons, and B mesons. 

The expected nuclear modification factors $\RAA$ were calculated 
in~\cite{adsw} exploring a large 
range in the medium density for central \mbox{Pb--Pb}
collisions at $\sqrtsNN=5.5~\tev$: $25<\hat{q}<100~\gev^2/\fm$. 
%Standard NLO perturbative predictions for the c and b 
%$\pt$-differential cross sections were used~\cite{hvqmnr}.
Figure~\ref{fig:lhc} (thick lines) 
shows the results for the heavy-to-light ratios 
of D and B mesons~\cite{adsw}, 
defined as the ratios of the nuclear
modification factors of D(B) mesons to that of light-flavour hadrons ($h$):
$R_{{\rm D(B)}/h}=\RAA^{{\rm D(B)}}/\RAA^h$. The effect of the mass 
is illustrated by artificially neglecting the  
mass dependence of parton energy loss (thin curves). 
The enhancement above unity that persists in the $m_{\rm c(b)}=0$ 
cases is mainly due to the
colour-charge dependence of energy loss,
since at LHC energy most of the light-flavour hadrons 
will originate from a gluon parent. The calculation
results indicate that, for D mesons,
the mass effect is small and 
limited to the region $\pt\lsim 10~\gev$, while for B
mesons a large enhancement can be expected up to $20~\gev$.
Therefore, the comparison of the high-$\pt$ suppression
for D mesons and for light-flavour hadrons would test the colour-charge 
dependence (quark parent vs. gluon parent) of parton energy loss,
while the comparison for B mesons and for light-flavour hadrons 
would test its mass dependence~\cite{adsw}.

\section{Heavy-flavour detection in ALICE}
\label{exp}

The ALICE experimental setup~\cite{alicePPR1} 
was designed so as to allow the detection
of ${\rm D}$ and ${\rm B}$ mesons in the high-multiplicity environment 
of central \PbPb~collisions at LHC energy, where up to several thousand 
charged particles might be produced per unit of rapidity. 
The heavy-flavour capability of the ALICE detector is provided by:
\begin{Itemize}
\item Tracking system; the Inner Tracking System (ITS), 
the Time Projection Chamber (TPC) and the Transition Radiation Detector (TRD),
embedded in a magnetic field of $0.5$~T, allow track reconstruction in 
the pseudorapidity range $-0.9<\eta<0.9$ 
with a momentum resolution better than
2\% for $\pt<20~\gev$ 
and a transverse impact parameter\footnote{The transverse impact parameter,
$d_0$, is defined as the distance of closest approach of the track to the 
interaction vertex, in the plane transverse to the beam direction.} 
resolution better than 
$60~\mum$ for $\pt>1~\gev$ 
(the two innermost layers of the ITS are equipped with silicon pixel 
detectors)\footnote{Note that, for pp collisions, the 
impact parameter resolution may be slightly worse, due to the 
larger transverse size of the beam at the ALICE interaction point.
This is taken into account in the studies presented here.}.
\item Particle identification system; charged hadrons are identified via 
$\dEdx$ in the TPC and in the ITS and via time-of-flight measurement in the 
Time Of Flight (TOF) detector; electrons are separated from charged 
pions in the dedicated TRD, and in the TPC; 
muons are identified in the forward muon 
spectrometer covering in acceptance the range $-4<\eta<-2.5$. 
\end{Itemize}

Detailed studies~\cite{alicePPR2}, 
based on full simulation of the detector and of the 
background sources, have shown that ALICE has a good potential to carry out
a rich heavy-flavour physics programme.  
Several analyses aimed at investigating quenching effects for 
c and b quarks are being prepared.
Here, we focus on the two most advanced studies in the central barrel:
exclusive reconstruction of charm particles, in the ${\rm D^0\to K^-\pi^+}$
decay channel, and inclusive measurement of beauty particles, 
in the semi-electronic decay channels ${\rm B}\to {\rm e}+X$.
Excellent performance is also expected for the measurement of beauty 
production at forward rapidity in the semi-muonic decay 
channels~\cite{alicePPR2,rachid}. In this context, the study of the 
single-inclusive muon distribution in the range $20\lsim\pt\lsim 50~\gev$ 
is a new promising tool to address energy loss effects for b quarks.
At LHC energy, single muons are dominated by decays of b quarks,
expected to strongly interact with the medium, 
for $3\lsim\pt\lsim 30~\gev$ and by decays of weakly-interacting,
thus `medium-blind', 
${\rm W^\pm}$ bosons for 
$\pt\gsim 30~\gev$~\cite{zaida}. Therefore, the position in $\pt$
of the crossing point between b-decay and W-decay muons should be 
sensitive to the in-medium energy loss of b quarks.

For all studies a multiplicity of $\dNdy=6000$
was assumed for central \PbPb~collisions\footnote{This value of the 
multiplicity can be taken as a conservative assumption, since 
extrapolations based on RHIC data predict $\dNdy=2000$--$3000$.}.
We report the results corresponding to the 
expected statistics collected by ALICE per LHC year: 
$10^7$ central (0--5\% $\sigma^{\rm inel}$) \PbPb~events at
$\mathcal{L}_{\rm Pb-Pb}=10^{27}~\cm^{-2}{\rm s}^{-1}$
and $10^9$ pp events at 
$\mathcal{L}_{\rm pp}^{\rm ALICE}=5\times 10^{30}~\cm^{-2}{\rm s}^{-1}$,
in the barrel detectors.

\subsection{Charm production: ${\rm D^0\to K^-\pi^+}$}

One of the most promising channels for open charm detection is the 
${\rm D^0 \to K^-\pi^+}$ decay which 
has a branching ratio (BR) of about $3.8\%$.
Based on next-to-leading order pQCD 
calculations~\cite{hvqmnr}, 
the expected production yields (${\rm BR}\times\d N/\d y$ at $y=0$) 
for ${\rm D^0}$ (+ ${\rm \overline{D^0}}$) 
mesons decaying in a ${\rm K^\mp\pi^\pm}$ pair 
in central 
Pb--Pb (0--$5\%~\sigma^{\rm inel}$) at $\sqrtsNN=5.5~{\rm TeV}$ and in pp 
collisions at $\sqrt{s}=14~{\rm TeV}$ are
$5.3\times 10^{-1}$ and $7.5\times 10^{-4}$ per event, 
respectively~\cite{alicePPR2}.

\begin{figure}[!t]
  \begin{center}
    \includegraphics[width=.5\textwidth]{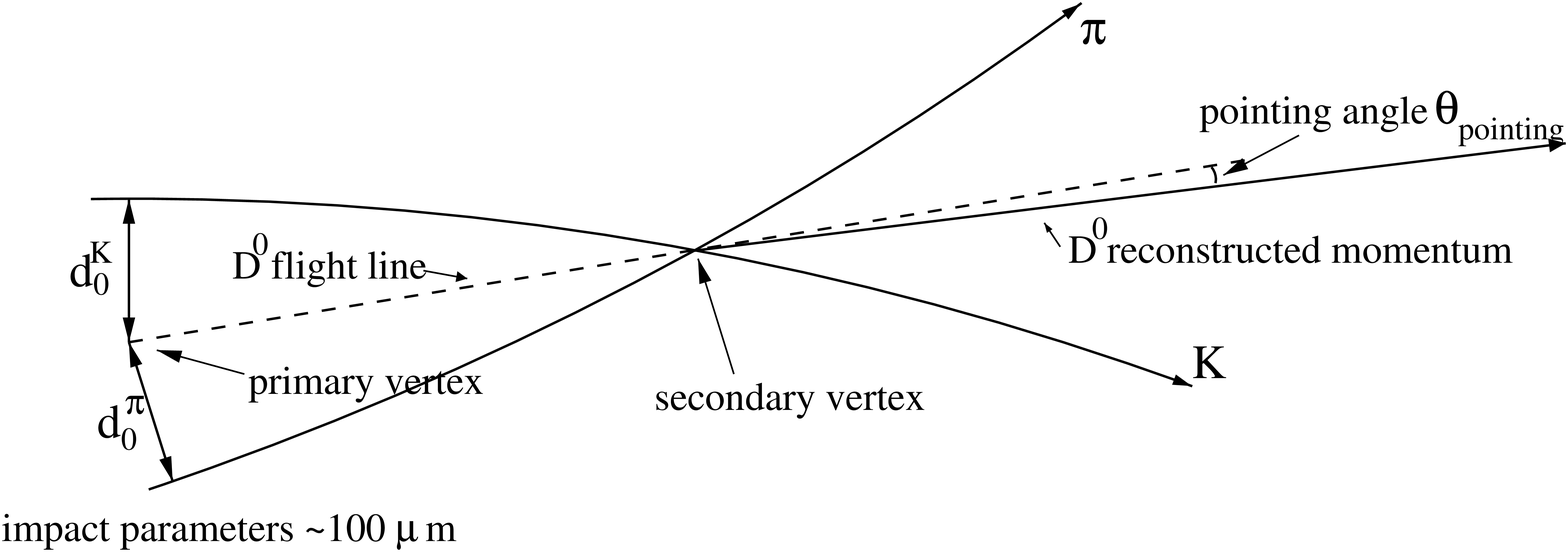}
    \caption{Schematic representation of the $\DtoKpi$ decay with
    the impact parameters ($d_0$) and the pointing angle
    ($\theta_{\rm pointing}$).} 
    \label{fig:D0sketch}
  \end{center}
\end{figure}  

\begin{figure}[!b]
  \begin{center}
    \includegraphics[width=.42\textwidth]{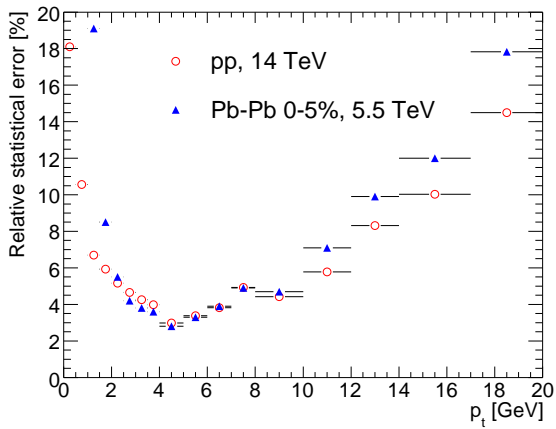}
    \caption{Expected relative statistical errors on the measurement 
             of the ${\rm D^0}$ production cross section, 
             in 0--5\% central Pb--Pb collisions 
             and in pp collisions.} 
    \label{fig:D0stat}
  \end{center}
\end{figure}

Figure~\ref{fig:D0sketch} 
shows a sketch of the decay: the main feature 
of this topology is the presence of two tracks with impact parameters 
$d_0\sim 100~\mum$. The detection strategy~\cite{D0jpg} to cope with
the large combinatorial background from the underlying event is based on:
\begin{Enumerate}
\item selection of displaced-vertex topologies, i.e. two tracks with 
large impact parameters
and small pointing angle $\theta_{\rm pointing}$ 
between the ${\rm D^0}$ momentum and flight-line
(see sketch in Fig.~\ref{fig:D0sketch});
\item identification of the K track in the TOF detector;
\item invariant-mass analysis.
\end{Enumerate}
This strategy was optimized separately for pp
and Pb--Pb collisions, as a 
function of the ${\rm D^0}$ transverse momentum~\cite{thesis,alicePPR2}. 
As shown in Fig.~\ref{fig:D0stat},
the accessible $\pt$ range is $1$--$20~\gev$ for Pb--Pb and 
$0.5$--$20~\gev$ for pp, 
with a statistical error better than 15--20\%. 
The systematic errors 
(acceptance and efficiency corrections, 
centrality selection for Pb--Pb) are expected to be smaller than 20\%. 
More details are given in Refs.~\cite{thesis,alicePPR2}.

\subsection{Beauty production: ${\rm B\to e}+X$}

The production of open beauty can be studied by detecting the 
semi-electronic decays of beauty hadrons, mostly B mesons. 
Such decays have a branching ratio of $\simeq 10\%$ 
(plus 10\% from cascade decays ${\rm b\to c \to e}$, that only populate 
the low-$\pt$ region in the electron spectrum).
The expected yields (${\rm BR}\times\d N/\d y$ at $y=0$)  
for ${\rm b\to e}+X$ plus ${\rm b}\to {\rm c}\,(\to {\rm e} +X)+X'$ 
in central Pb--Pb ($0$--$5\%~\sigma^{\rm inel}$) at $\sqrtsNN=5.5~{\rm TeV}$
and in in pp collisions at $\sqrt{s}=14~{\rm TeV}$ 
are $1.8\times 10^{-1}$  and $2.8\times 10^{-4}$ per event, 
respectively~\cite{alicePPR2}.

\begin{figure}[!b]
  \begin{center}
    \includegraphics[width=.42\textwidth]{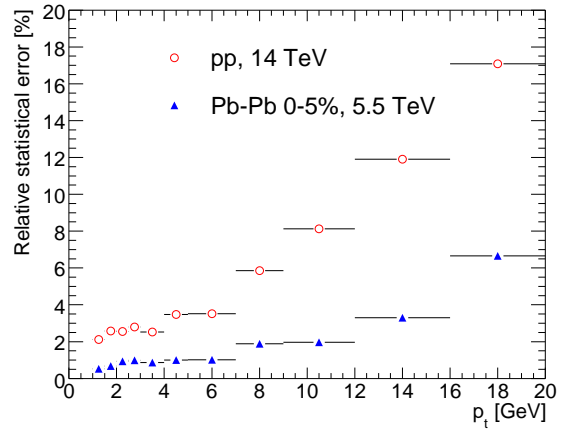}
    \caption{Expected relative statistical errors on the measurement 
             of the production cross section of B-decay electrons, 
             in 0--5\% central Pb--Pb collisions 
             and in pp collisions.} 
    \label{fig:eBstat}
  \end{center}
\end{figure}  

The main sources of background electrons are: (a) decays of D mesons; 
(b) neutral pion Dalitz decays $\pi^0\to \gamma {\rm e^+e^-}$ 
and decays of light mesons (e.g.\,$\rho$ and $\omega$);
(c) conversions of photons in the beam pipe or in the inner detector 
layer and (d) pions misidentified as electrons. 
Given that electrons from beauty have average 
impact parameter $d_0\simeq 500~\mum$
and a hard momentum spectrum, it is possible to 
obtain a high-purity sample with a strategy that relies on:
\begin{Enumerate}
\item electron identification with a combined $\dEdx$ (TPC) and transition
radiation selection, which is expected to reduce the pion contamination 
by a factor $10^4$;
\item impact parameter cut to reject misidentified $\pi^\pm$ and $\rm e^{\pm}$
from Dalitz decays and $\gamma$ conversions 
(the latter have small impact parameter for $\pt\gsim 1~\gev$);
\item $\pt$ cut to reject electrons from charm decays. 
\end{Enumerate}
As an example, with $d_0>200~\mum$ and $\pt>2~\gev$, the expected statistics
of electrons from b decays is $8\times 10^4$ for $10^7$ central 
Pb--Pb events, allowing the measurement of electron-level 
$\pt$-dif\-fe\-ren\-tial 
cross section in the range $2<\pt<20~\gev$. 
The residual contamination of about 10\% 
of electrons from prompt charm decays, from misidentified charged pions
and $\gamma$-conversion electrons 
can be evaluated and subtracted using a Monte Carlo simulation tuned 
to reproduce the measured cross sections for pions and 
$\rm D^0$ mesons.
In Fig.~\ref{fig:eBstat} we show the expected relative statistical 
errors on the measurement of the cross section of electrons from beauty 
decays.
A Monte-Carlo-based procedure can then be used to compute,
from the electron-level cross section, the B-level cross section 
$\d\sigma^{\rm B}(\pt>\pt^{\rm min})/\d y$~\cite{alicePPR2}. 
The covered range is $2<\pt^{\rm min}<30~\gev$.

\section{Sensitivity to energy loss}
\label{sensitivity}

We investigated the possibility of using 
the described charm and beauty measurements 
to study the
dependences of parton energy loss. This study 
could be carried out by measuring:
\begin{Itemize} 
\item the nuclear modification factor of 
D mesons as a function of transverse momentum, $\RAA^{\rm D}(\pt)$;
\item the nuclear modification factor of 
b-decay electrons as a function of transverse momentum, 
$\RAA^{\rm e\,from\,B}(\pt)$;
\item the heavy-to-light ratio, $\RDh(\pt)$, defined
as the ratio of the nuclear modification factors of 
D mesons and of charged light-flavoured hadrons.
\end{Itemize}
We compare the expected experimental errors 
on these observables to recent theoretical predictions 
parton energy loss~\cite{adsw}.

\begin{figure}[!b]
  \begin{center}
    \includegraphics[width=.47\textwidth]{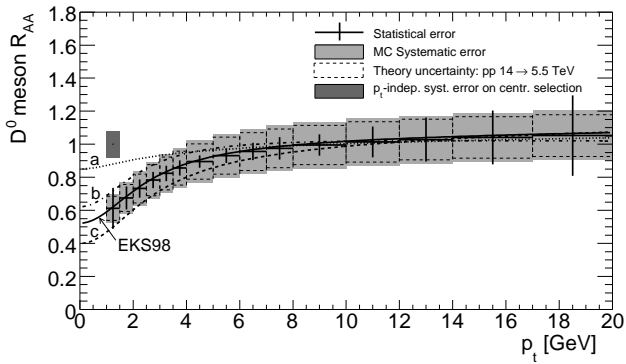}
    \caption{Illustration of the experimental uncertainties
             on the $\RAA$ of $\Dz$ mesons (no energy loss).} 
    \label{fig:RAAshad}
  \end{center}
\end{figure}

The expected performance for the measurement of 
the nuclear modification factor for $\Dz$ mesons is reported in 
Fig.~\ref{fig:RAAshad}. Only nuclear shadowing and 
parton intrinsic 
trans\-ver\-se-momentum broadening are included (no energy loss).
The reported 
statistical (bars) and systematic (shaded areas) errors are obtained combining 
the errors for \mbox{Pb--Pb} and pp collisions and 
assuming that the contributions due to cross section normalization, 
feed-down from beauty decays and, partially, acceptance/efficiency corrections
will cancel out in the ratio. An expected 
uncertainty of about 12\%~\cite{alicePPR2} introduced in the
extrapolation of the pp results from 14~TeV to 5.5~TeV by pQCD
is also included. 

The effect of shadowing, 
included via the EKS98 pa\-ra\-me\-tri\-sa\-tion~\cite{EKS98},
is visible as a suppression of $\RAA$ at low transverse momenta, 
corresponding to small Bjorken $x$.
The effect is negligible for 
$\pt>6$--$7~\gev$. Since there is significant uncertainty on the 
magnitude of shadowing in the low-$x$ region, we studied the effect of such
uncertainty on $\RAA$ by varying the nuclear modification of 
parton distribution functions. 
Also in the case of shadowing 50\% stronger than in EKS98 (curve labelled `c'),
we find $\RAA>0.95$ for $\pt>8~\gev$. Under these assumptions, for 
$\pt>8~\gev$ only parton energy loss is expected to affect 
the nuclear modification factor of D mesons.

\begin{figure}[!t]
  \begin{center}
    \includegraphics[width=0.5\textwidth]{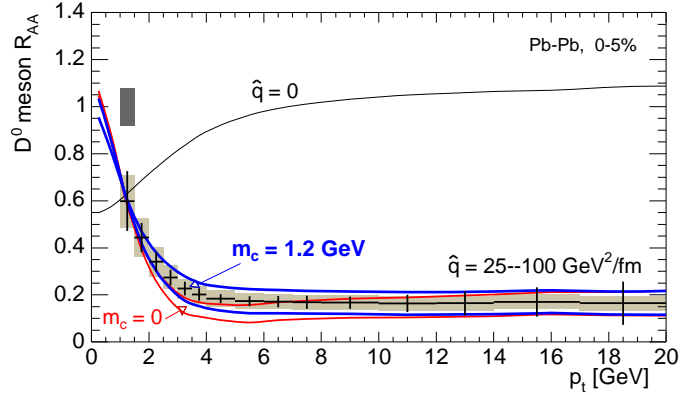}
    \caption{Nuclear modification factor for $\Dz$ mesons with shadowing.
             The two bands represent theoretical predictions with 
             and without the effect of the charm mass in the energy loss
             for the medium transport coefficient range 
             $\hat{q}=25$--$100~\gev^2/\fm$~\cite{adsw}.
             Errors corresponding to 
             the centre of the 
             prediction band for $m_{\rm c}=1.2~\gev$ are shown: 
             bars = statistical, 
             shaded area = systematic.
             The normalization error is shown by the box at $\RAA=1$.} 
    \label{fig:RAAD0}
  \end{center}
\end{figure}

\begin{figure}[!t]
  \begin{center}
    \includegraphics[width=0.5\textwidth]{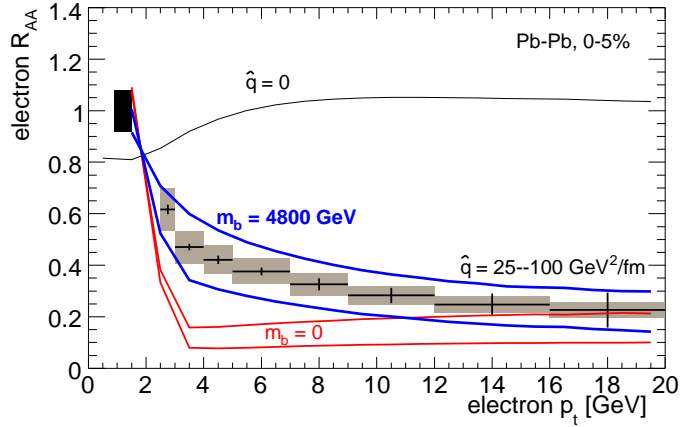}
    \caption{Nuclear modification factor for b-decay electrons with shadowing.
             The two bands represent theoretical predictions with 
             and without the effect of the beauty mass in the energy loss
             for the medium transport coefficient range 
             $\hat{q}=25$--$100~\gev^2/\fm$~\cite{adsw}.
             Errors are defined as for Fig.~\ref{fig:RAAD0}.} 
    \label{fig:RAAeB}
  \end{center}
\end{figure}

\begin{figure}[!t]
  \begin{center}
    \includegraphics[width=0.5\textwidth]{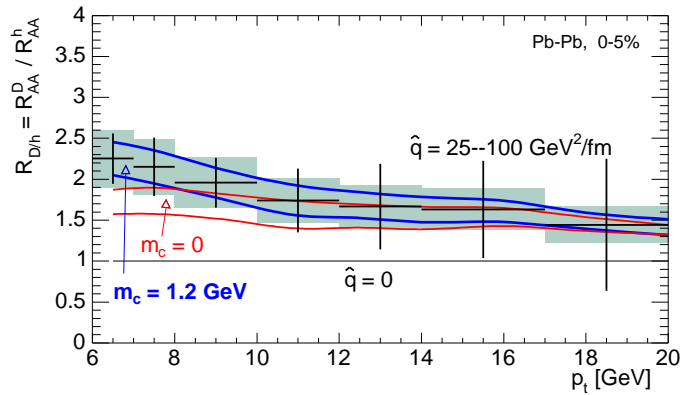}
    \caption{Ratio of the nuclear modification factors for $\Dz$ mesons 
             and for charged hadrons. 
             Errors corresponding to the centre of the prediction band   
             for $m_{\rm c}=1.2~\gev$ are shown: bars = statistical, 
             shaded area = systematic.} 
    \label{fig:RDh}
  \end{center}
\end{figure}

Figure~\ref{fig:RAAD0} presents the predicted~\cite{adsw} 
nuclear modification factor 
without ($\hat{q}=0$) and with energy loss (the bands correspond
to the range $25<\hat{q}<100~\gev^2/\fm$). 
The effect of the charm mass on energy loss is included for 
the thick-line band ($m_{\rm c}=1.2~\gev$) and not included 
for the thin-line band
($m_{\rm c}=0$). The small difference between the two bands 
indicates that, with respect to energy loss, 
charm behaves essentially as a massless quark. 
%The rapid increase of 
%$\RAA$ as $\pt\to 0$ is due to the fact that, in the calculation adopted 
%in Ref.~\cite{adsw}, the charm quarks that lose most of their initial energy 
%in the medium are assumed to thermalize and give a component with a 
%steeply falling spectrum at low $\pt$.
The estimated uncertainties for the measurement of $\RAA^{\rm D}$ are
reported here for the case {\it with} energy loss. The bars represent the 
statistical errors, while the shaded areas represent the quadratic sum 
of the systematic error from Monte Carlo corrections ($\approx 15\%$) 
and that from 
the $\sqrt{s}$ extrapolation of the pp measurement from 14~TeV to 5.5~TeV
($\approx 12\%$).
Owing to the predicted 
suppression of about a factor 5 for $\pt\gsim 5~\gev$, the relative
statistical errors in \mbox{Pb--Pb} are larger by more than a factor 2, 
with respect to the case of no suppression, and they become the 
dominant contribution to the statistical error on $\RAA^{\rm D}$. 

The expected performance for the measurement of the nuclear
modification factor of electrons from B-meson decays is shown in 
Fig.~\ref{fig:RAAeB}, along with the predicted suppression 
with and without the effect of the beauty mass in the energy loss.
Contrary to the case of charm, the suppression is strongly 
reduced in the range $2\lsim\pt\lsim 15~\gev$ 
due the large value of the b mass. 
As the charm case,
the bars represent the 
statistical errors, while the shaded areas represent the quadratic sum 
of the systematic error from Monte Carlo corrections ($\approx 15\%$) 
and that from 
the $\sqrt{s}$ extrapolation of the pp measurement from 14~TeV to 5.5~TeV
($\approx 8\%$).

As discussed in Section~\ref{eloss},
the comparison of the high-$\pt$ suppression 
of charm-quark-originated mesons and 
gluon-originated hadrons may be the tool best suited to 
single out the predicted colour-charge dependence of QCD energy loss.
The ALICE sensitivity to the heavy-to-light ratio $\RDh$ in the 
range $5<\pt<20~\gev$ is 
presented in Fig.~\ref{fig:RDh}. Like for the case of $\RAA^{\rm D}$, 
the two bands correspond to including or not including the effect of the
c-quark mass for a medium transport coefficient in the range 
$25$--$100~\gev^2/\fm$. For $10<\pt<20~\gev$, the two bands 
coincide and predict $\RDh\approx 1.5$, i.e., about 50\% smaller 
suppression for D mesons relative to light-flavoured hadrons.
Many of the systematic uncertainties on $\RDh$ 
cancel out (centrality selection and, partially, acceptance/efficiency 
corrections and energy extrapolation by pQCD)
since $\RDh$ is essentially a double ratio. 
The residual systematic 
error is estimated to be of about 15\%.
We assumed the statistical error on $\RAA^h$ to be negligible with 
respect to that on $\RAA^{\rm D}$ for $\pt<20~\gev$. 
The resulting statistical errors on $\RDh$ are quite large for 
$\pt\gsim 15~\gev$. However, at lower momenta, the measurement of the 
compatibility (or incompatibility) of $\RDh$ with unity 
appears to be feasible. 
A procedure to use the electrons from beauty decays to estimate
the heavy-to-light ratio of B mesons (lower panel
of Fig.~\ref{fig:lhc}) is currently under study.

\section{Conclusions}
\label{conclusions}

Heavy-quark quenching studies have become one of the 
most intriguing topics in heavy-ion physics, with the 
observation at RHIC of a large suppression for heavy-flavour
decay electrons, which is at present not clearly 
understood within parton energy loss models.

At LHC energy, charm and beauty production cross sections 
are expected to be larger by factors of approximately 10 and 100,
respectively, with respect to RHIC energy.
The ALICE experiment will be equipped with high-resolution silicon 
vertex detectors, allowing direct and precise
measurements of the main observables that are suggested to be 
sensitive to the colour-charge and mass 
dependences of parton energy loss.

\end{document}